%% file: main.tex
\documentclass[10pt,conference]{IEEEtran}

\usepackage[utf8]{inputenc}
\usepackage[english]{babel}
\usepackage{booktabs}
\usepackage{graphicx}
\usepackage{listings}
\usepackage{multirow}
\usepackage{array}
\usepackage{soul}
\usepackage{url}
\usepackage{xcolor}

\usepackage{mathtools}

\DeclarePairedDelimiter{\floor}{\lfloor}{\rfloor}

\makeatletter
\newcommand{\linebreakand}{%
  \end{@IEEEauthorhalign}
  \hfill\mbox{}\par
  \mbox{}\hfill\begin{@IEEEauthorhalign}
}
\makeatother

\author{
 \IEEEauthorblockN{Yaroslav Golubev,\IEEEauthorrefmark{1} Viktor Poletansky,\IEEEauthorrefmark{2} Nikita Povarov,\IEEEauthorrefmark{4} Timofey Bryksin\IEEEauthorrefmark{1}\IEEEauthorrefmark{2}}
    \IEEEauthorblockA{\IEEEauthorrefmark{4}\textit{JetBrains}, \IEEEauthorrefmark{1}\textit{JetBrains Research},  \IEEEauthorrefmark{2}\textit{Saint Petersburg State University}}
    \IEEEauthorblockA{yaroslav.golubev@jetbrains.com, poletansky@gmail.com, nikita.povarov@jetbrains.com, timofey.bryksin@jetbrains.com}
}

\title{Multi-threshold token-based code clone detection}

\begin{document}

\maketitle

\begin{abstract}

Clone detection plays an important role in software engineering. Finding clones within a single project introduces possible refactoring opportunities, and between different projects it could be used for detecting code reuse or possible licensing violations.

In this paper, we propose a modification to bag-of-tokens based clone detection that allows detecting more clone pairs of greater diversity without losing precision by implementing a multi-threshold search, i.e. conducting the search several times, aimed at different groups of clones. To combat the increase in operation time that this approach brings about, we propose an optimization that allows to significantly decrease the overlap in detected clones between the searches. 

We evaluate the method for two different popular clone detection tools on two datasets of different sizes. The implementation of the technique allows to increase the number of detected clones by 40.5--56.6\% for different datasets. BigCloneBench evaluation also shows that the recall of detecting Strongly Type-3 clones increases from 37.5\% to 59.6\%.

\end{abstract}

\input{sections/01-introduction}
\input{sections/02-background}
\input{sections/03-methodology}
\input{sections/04-parameters}
\input{sections/05-evaluation}
\input{sections/06-threats}
\input{sections/07-conclusion}

\bibliographystyle{ieeetran}
\bibliography{cites}

\end{document}

%% file: sections/01-introduction.tex
\section{Introduction}

When two pieces of code are the same or are similar to each other, they are called \textit{clones}. Several studies have shown that a significant part of modern software code consists of clones~\cite{roy2008empirical, dejavu}. Detection of such clones is an important problem in software engineering that has been significantly developed in the last decades.
Research has shown that maintaining cloned bugs within a project presents serious challenges, and at the same time, providing the users with the information about clones in the code can significantly increase the efficiency of bug detection and fixing~\cite{chatterji2013effects, mondal2019investigating}.

On the other hand, inter-project clones are relevant for detecting plagiarism and license violations~\cite{german2009code, golubev2020study} or identifying potential library candidates from the most used pieces of code~\cite{libraries}. Large-scale inter-project clone detection often relies on \textit{token-based} tools~\cite{dejavu, golubev2020study}, which has to do with the fact that such approaches are very fast and can be expedited even more with various heuristics. Such scalability is very important for large datasets. 

One of the problems with token-based clone detection, and clone detection in general, is that it is highly configurable and finding the best search parameters is a difficult task~\cite{wang2013searching}.

The goal of this paper is to alleviate this problem for some of the token-based clone detection tools by considering the relationship between the size of code fragments and their similarity. Our idea stems from an observation that code fragments of different sizes require different similarity thresholds -- larger code blocks have more room for modification that would still leave them similar enough to be considered clones. Therefore, we propose to detect clones with a \textit{parametric curve}, i.e. a set of different parameter configurations (similarity threshold and the length of pieces of code in tokens) aimed at different groups of clones.

In this paper, we evaluate the idea for blocks-based bag-of-tokens clone detection tools. We implemented the proposed approach and estimated the necessary parameter configurations for two prominent bag-of-tokens clone detection tools, SourcererCC~\cite{SourcererCC} and CloneWorks~\cite{svajlenko2017cloneworks}. We evaluated the method using the clone detection tools on two datasets consisting of four prominent open-source repositories. The tools showed an increase of 40.5--56.6\% in the number of detected clones on different datasets. We also studied the recall of the method by running CloneWorks with and without it on a popular dataset BigCloneBench~\cite{svajlenko2016bigcloneeval}, which demonstrated an increase in Strongly Type-3 Recall from 37.5\% to 59.6\%.

Finally, to combat the increase in the operation time, we propose an optimization to this technique that allows to decrease the overlap in discovered clones. The optimization consists in estimating the most extreme cases of possible block sizes in a clone pair and excluding pairs that are guaranteed to be discovered in another search. For example, when detecting clones with SourcererCC on a larger of two datasets, the optimization decreased the operation time of the technique by 12.7\%, making a total of eight runs only 2.5 times slower than performing a clone detection with a single parameter configuration. 

The proposed approach was used in our previous large-scale study to detect license violations in Java code on GitHub~\cite{golubev2020study}, where it demonstrated its usefulness.

%% file: sections/02-background.tex
\section{Background}\label{related}

Code clone detection is a mature field, and there are different tools aimed at different types of clones. Code can be represented as lines of code, tokens, abstract syntax tree, program dependency graph, etc.~\cite{belloncomparison}. A recent review of clone detection approaches~\cite{ain2019recent} lists six main categories of techniques: textual, token-based, tree-based, metric-based, semantic, and hybrid.

Token-based techniques, despite being on the simpler end of the spectrum, are still very popular and widely used. Their main disadvantage is the inability to detect semantic clones, however, not every clone detection task is aimed at detecting semantic clones. The main reason token-based clone detection tools are widespread is their good scaling, which makes them uniquely qualified for extremely large datasets of millions of lines of code. Popular token-based clone detection tools include CCFinder~\cite{kamiya2002ccfinder}, SourcererCC~\cite{SourcererCC}, CloneWorks~\cite{svajlenko2017cloneworks}.

However, one of the problems with token-based clone detectors (and with clone detectors in general) is the necessity to pick specific search configurations. The tools are highly configurable and typically come with some default settings that developers tested and evaluated but that are not the best for all the cases and datasets.

To combat this problem, a very interesting study was conducted by Wang et al.~\cite{wang2013searching}. As their motivation, the authors looked at 275 software clones papers and found that of those that had an empirical study, 61\% commented on the problems arising from the choice of configuration. The authors then developed a pipeline based on a genetic algorithm that searches for optimal parameters from the standpoint of the agreement of several clone detection tools and presented their own application called EvaClone based on it. The authors conducted an empirical evaluation of EvaClone and discovered that their method can produce better parameter configurations both for specific projects and in general. They also demonstrated that default tool configurations are often imperfect. A replication study of EvaClone conducted by Ragkhitwetsagul et al.~\cite{ragkhitwetsagul2016searching} also showed that the optimized parameters produce better results than the default ones. The authors also pointed out certain weaknesses of the approach's fitness function. This problem was also discussed by Gauci~\cite{gauci2015smelling}.

Ragkhitwetsagul et al.~\cite{ragkhitwetsagul2018comparison} also compared various clone detection techniques in specific scenarios, including pervasive code modifications and boilerplate code. After a thorough evaluation of various configurations, the authors concluded that default parameters do not show the tools' best results.

Finally, a very unique research was carried out by Keivanloo et al.~\cite{keivanloo2015threshold}. The authors proposed the idea that a large amount of heterogeneous data should not be searched using a single threshold at all. In a motivating example, the authors showed that different functionalities of a diverse project can vary differently and therefore require different similarity thresholds. The authors then proposed an approach that uses \textit{k}-means clustering algorithm to divide the target code blocks (Java methods), and distinguishes between true and false clone pairs using the clusters. Overall, the authors showed that such threshold-free clone detection shows the F-measure to be 12\% better than for a single threshold.

While Keivanloo et al. suggest to consider different thresholds for different functionalities, we propose to consider different thresholds for different sizes of code blocks. In our study, we work with two parameters --- minimum size in tokens and similarity threshold --- and apply them to bag-of-tokens clone detection.

%% file: sections/03-methodology.tex
\section{Methodology}\label{methodology}

\subsection{Possible clones space and parametric curve}\label{parametric}

\begin{figure}[h]
  \centering
  \vspace{-0.1cm}
  \includegraphics[width=3.45in]{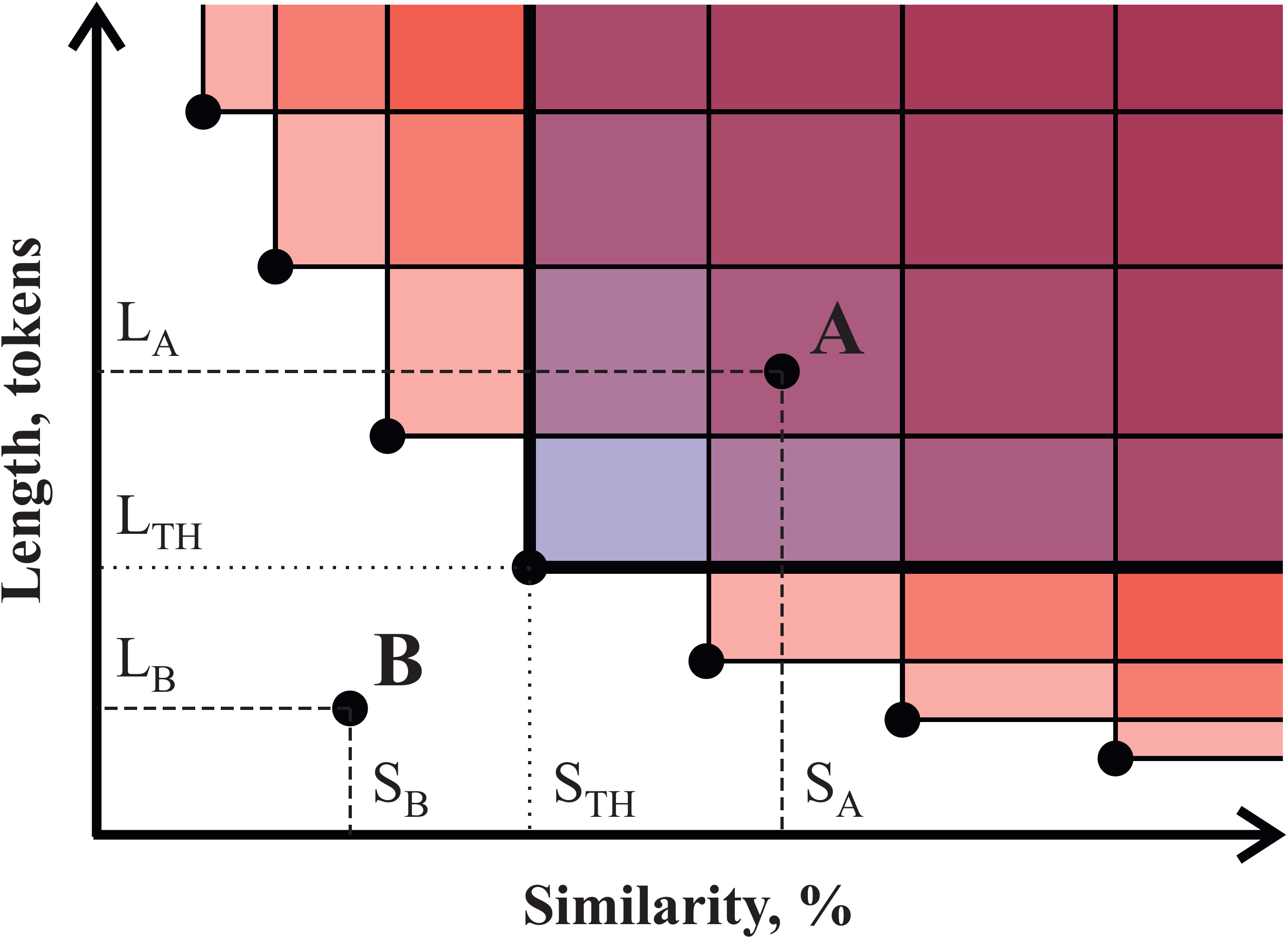}
  \caption{Possible clones space. The highlighted area in bold line represents possible search results with a single parameter. All together, different areas represent possible search results with a parametric curve.}
  \label{point}
  \vspace{-0.1cm}
\end{figure}

The idea of the proposed modification to bag-of-tokens based clone detection is demonstrated and explained in Figure~\ref{point}. Every pair of code blocks can be expressed as a pair of two numbers: its length in tokens (the length of the smaller of two blocks) and its similarity (which in the bag-of-tokens model is the number of tokens they share divided by the total number of tokens in the larger block). These numbers can be used as coordinates, and the pair A \( (S_{A},L_{A}) \) in Figure~\ref{point} contains blocks that are larger and more similar than those in pair B \( (S_{B},L_{B}) \). From this standpoint, choosing the parameters for running token-based clone detection (Similarity Threshold and Lower Token Length Threshold) means choosing the area of the plane that a researcher considers to represent valid clones they want to detect. 

When a researcher chooses the parameters, only the pairs that are larger and more similar than these parameters are considered to be clones. In the example of Figure~\ref{point}, the pair of parameters \( (S_{TH},L_{TH}+) \)\footnote{From here on out we will use such notation for search configurations, the ``+'' sign indicates that there is no Upper Token Length Threshold.} yielded the area highlighted in blue in bold borders. With such search configuration, the pair A would be detected as clones, while the pair B would not, because the blocks in pair B would not be considered large enough and similar enough. 

However, valid clones are not spread evenly on this plane. Naturally, more of them lean to the right side where the similarity is higher. What is even more important, we hypothesize that larger blocks require less similarity to be valid (see highlighted areas in Figure~\ref{point}). The reason for this lies in sub-block similarities, in other words, larger blocks have more room for modification so that the resulting blocks are still similar enough to be considered clones. As an extreme example, consider two methods of 40 lines each, where 20 lines are the same. Even though the similarity is only 50\% (less than researchers usually set as a threshold), such a pair would still be of interest both for refactorings or plagiarism detection. 

If that is the case, it is possible to conduct several instances of clone detection with different parameter configurations aimed at different groups of clones and merge the results together to detect more clones of a more diverse nature, which might be of special interest to the task of detecting code reuse between different projects. We call this set of different parameter configurations a parametric curve (see points in Figure~\ref{point}). This method produces a large overlap of the detected clone pairs and can therefore be considered as converting temporal or computational resources into the number of detected pairs, which is well suited for large-scale studies that require a single case of clone detection and value fullness over cost-effectiveness, such as detecting possible violations or studying the developing practices in a large corpus of projects. Different searches can also be run in parallel.

\subsection{Optimization of the search process}\label{sec:optimizing}

The main disadvantage of the described approach is the significant amount of extra operating time or resources spent on finding the same clone pairs in different runs of clone detection (we refer to them as \textit{search instances}), which is visualized as an overlap of rectangular areas in Figure~\ref{point} -- the clone pair A is detected three times. However, if the similarity function is simply the fraction of the shared tokens, this can be optimized. 

The optimization employs the fact that you can set not only the Lower Token Length Threshold but also the Upper one. To calculate the Upper Token Length Threshold for each search instance, one needs to account for the most extreme possible difference in block length. Let us say that parameter pairs \((75, 40+)\), \((70, 60+)\) are adjacent on the curve, and let us calculate the upper bound for the first one. The largest block that the second instance will not process is 59 blocks. The largest block that can form a valid clone pair with such a block would have to include \textit{all} of its tokens and have as many different tokens as the Similarity Threshold would allow. That means that the Upper Token Length Threshold is equal to \(\*{\floor{\frac{59}{0.75}}} = 78\). At Similarity Threshold of 75\%, any block larger than that \textit{can not have} valid clone pairs smaller than 59 tokens, meaning that a pair of blocks of sizes 59 and 79 does not need to be detected because it can \textit{never} be clones with our parameters. As for a pair of blocks of sizes 60 and 79, it does not need to be detected because if it is a valid clone, it will be detected in the instance with parameters \((70, 60+)\).

Therefore, in order not to miss any pairs and at the same time not to do any unnecessary work, one must run this search with parameters (75, 40--78) and (70, 60+), which will account for the most extreme difference in blocks' size. Naturally, the Upper Token Length Threshold of the second instance can be calculated the same way, taking into account the \textit{next} search instance, and so on.

In general, the Upper Token Length Threshold for the \textit{i}-th instance is calculated as:

\begin{equation} \label{eq:1}
UTLT_{i} = \floor*{\frac{LTLT_{i+1} - 1}{ST_{i}}}
\end{equation}

where \(LTLT_{i+1}\) is the Lower Token Length Threshold of the next (less similar) search instance and \(ST_{i}\) is the \textit{i}-th Similarity Threshold. 

This calculation is carried out from the most similar search instance (i.e. the search instance with the largest Similarity Threshold) downwards. For the least similar search instance, there is no next one, so it has no Upper Token Length Threshold and therefore cannot be optimized in this fashion. 

%% file: sections/04-parameters.tex
\section{Estimating the parameters}\label{estimating}

\begin{table*}
  \centering
\begin{tabular}{@{}ccccccccc@{}}
\toprule
\multirow{2}{*}{\textbf{Tool}}        & \multirow{2}{*}{\textbf{Dataset}} & \multicolumn{3}{c}{\textbf{Detected clones}}            & \multicolumn{4}{c}{\textbf{Operation time}}                             \\ \cmidrule(l){3-5} \cmidrule(l){6-9}
                             &                          & \textbf{BC} & \textbf{PC} & \textbf{Delta}  & \textbf{BC} & \textbf{PC Raw}   & \textbf{PC Optimized} & \textbf{Delta (PC Optimized / BC)} \\ \midrule \vspace{0.1cm}
\multirow{2}{*}{\textbf{SourcererCC}} & \textbf{Smaller}                  & 35,836             & 50,356           & 40.5\% & 3 min 12 sec       & 19 min 12 sec & 18 min 38 sec     & 482\% \\ 
                             & \textbf{Larger}                   & 136,591            & 193,411          & 41.6\% & 10 min 59 sec      & 32 min 12 sec & 28 min 07 sec     & 156\% \\ \midrule \vspace{0.1cm}
\multirow{2}{*}{\textbf{CloneWorks}}  & \textbf{Smaller}                  & 20,172             & 31,584           & 56.6\% & 26 sec             & 38 sec        & 34 sec            & 31\%  \\ 
                             & \textbf{Larger}                   & 27,309             & 40,955           & 50\%   & 3 min 58 sec       & 4 min 29 sec  & 4 min 20 sec      & 9\%   \\ \bottomrule \vspace{0.01cm}
\end{tabular}
  \caption{The results of the projects-based evaluation. BC stands for Best single-threshold configuration, PC -- Parametric curve.}
  \label{results}
  \vspace{-0.5cm}
\end{table*}

We have performed a manual estimation of the necessary configurations for two prominent clone detection tools that use the bag-of-tokens model, SourcererCC~\cite{SourcererCC} and CloneWorks~\cite{svajlenko2017cloneworks}. Both of these tools support several languages of input code, in our study we have chosen Java.

\subsection{The experiment}

The parameter pairs of the discussed parametric curve are inherently unknown and must be determined experimentally. To determine them, we have manually evaluated clones found within a dataset of three projects: \textit{Spring Framework}, \textit{Guava}, and \textit{OpenJDK}. These projects have more than a thousand stars on GitHub and were chosen for their maturity. The clones from SourcererCC and CloneWorks were studied separately.

The dataset was tokenized with a block-level of granularity, and we chose the Lower Token Length Threshold to be at least 19 tokens. This value was obtained in preliminary experiments with manual evaluation, other researchers used similar values: for example, Saini et al.~\cite{saini2018cloned} use the threshold of 25 tokens. 

With this threshold of 19 tokens, we carried out 31 runs of clone detection for each tool with Similarity Threshold values in the range from 50\% to 80\%, with a step of 1\%. That range was chosen because it contains a majority of values that previous researchers considered to be optimal. After that, the first and the second authors labeled the pairs as either true positive or false positive as follows.

For the largest Similarity Threshold (80\%), the authors labeled the largest pairs of clones, and then repeated the process while gradually decreasing the Token Length of the blocks in the pair and picking samples. The process was stopped when the precision fell down to 90\%, the current Token Length of the last pair was established as a Lower Token Length Threshold for this Similarity Threshold. This makes it so that the precision above the curve stays above 90\%, which we consider sufficient for large-scale clone detection.

The previous step was repeated for the next, lower Similarity Threshold, excluding all the previously found pairs. Since the run on any given Similarity Threshold includes all the results from the more similar runs with the same Token Length Thresholds, we were only interested in the newly found pairs. 

Once we conducted this process for every studied value of the Similarity Threshold, we drew up the results for each tool in a single table. If two adjacent Similarity Thresholds corresponded to the same value of Lower Token Length Threshold, only the smaller value of Similarity Threshold was kept, because, as previously mentioned, it contained all the results from the higher value (see clone detection space in Figure~\ref{point}).

As a final step, we calculated the Upper Token Length Threshold using Equation~\ref{eq:1}.

\begin{table}
\centering
  \caption{Parametric curve configurations. ST is Similarity Threshold, LTLT and UTLT are Lower and Upper Token Length thresholds, respectively.}
  \label{table:optimized}
  \begin{tabular}{ccccccccc}
    \multicolumn{9}{c}{\textbf{SourcererCC}}\vspace{0.1cm}\\
    \toprule
    \textbf{ST} & 75 & 73 & 71 & 70 & 65 & 60 & 55 & 50\\
    \midrule \vspace{0.1cm}
    \textbf{LTLT} & 19 & 24 & 34 & 36 & 56 & 65 & 144 & 215\\
    \textbf{UTLT} & 30 & 45 & 49 & 78 & 98 & 238 & 389 & \(\infty\)\\
    \bottomrule\vspace{0.01cm}

\end{tabular}
  \begin{tabular}{ccccccccc}
    \multicolumn{9}{c}{\textbf{CloneWorks}}\vspace{0.1cm}\\
    \toprule
    \textbf{ST} & 77 & 75 & 72 & 71 & 70 & 65 & 60 & 55\\
    \midrule \vspace{0.1cm}
    \textbf{LTLT} & 19 & 22 & 24 & 30 & 35 & 50 & 72 & 140\\
    \textbf{UTLT} & 27 & 30 & 40 & 47 & 70 & 109 & 231 & \(\infty\)\\
  \bottomrule
\end{tabular}
\vspace{-0.2cm}
\end{table}

\subsection{Parameter pairs}\label{sec:parampairs}

The resulting parametric curve configurations for both tools are presented in Table~\ref{table:optimized}. For SourcererCC, tokenization was performed in blocks mode. For CloneWorks, the tokenization was performed for functions, the type of tokenization was chosen to be \textit{type3token} as the most similar one to SourcererCC, however, different implementations of tokenization in tools can lead to different results of clone detection. The resulting parameter pairs strongly correlate to the ones of SourcererCC, indicating that different bag-of-tokens based clone detectors behave similarly in this regard.

We have also noticed that our hypothesis stated in Section~\ref{parametric} is true: the larger the blocks, the lesser similarity they require to still be considered clones. At relatively low similarity thresholds, a detected clone pair of small blocks can sometimes present a shuffle of keywords and identifiers, while large blocks still look more similar, albeit with large modifications.

%% file: sections/05-evaluation.tex
\section{Evaluation}\label{sec:results}

\subsection{Datasets}

To evaluate the increase in the number of detected clone pairs with the proposed method, as well as to capture the effects of the possible optimization, we performed an evaluation on two datasets of different sizes to better capture the efficiency of the possible scaling. 

For the first, \textit{Smaller} dataset, we chose three other prominent Java repositories: \textit{Mockito}, \textit{Glide}, and \textit{RxJava} because of their popularity and their relatively small size. SourcererCC tokenization reveals just 44,864 unique blocks of code in this dataset. 

To test the scalability of the optimization and also to run it in a completely different setting, we have used a second, \textit{Larger}, dataset an order of magnitude larger than the previous one, which consisted of a single large project --- \textit{IntelliJ IDEA Community Edition}. SourcererCC tokenization parsed this repository into 321,478 blocks of code. Not only would the operation time be different here, but the clones in a single large system can also have an entirely different structure.

Finally, to closer analyze the obtained extra clones we have also run CloneWorks on the popular benchmark BigCloneBench~\cite{svajlenko2016bigcloneeval} with and without the parametric curve. BigCloneBench is an established and mature dataset that also splits the clone pairs by types, which can allow us to understand what clones are added. We used BigCloneBench evaluation with the same minimum of 19 tokens, the replication package for this experiment is available online.\footnote{Replication package for BCB: \url{https://doi.org/10.5281/zenodo.4279693}}

\subsection{Projects}

The overall results for the two projects-based datasets are presented in Table~\ref{results}, however, detailed tables with the results for each configuration of each tool on each dataset are available online.\footnote{Detailed results: \url{https://doi.org/10.5281/zenodo.4274230}} One can make the following observations from these results.

In all four cases, our approach allows us to detect significantly more clones, increasing the number of detected clones by 40.5--56.6\%.

The increase in the operation time strongly depends on the operating process of the tool. SourcererCC has a slower clone detection stage, therefore, running it eight times results in a significant increase in time. On the other hand, CloneWorks has a longer tokenization stage and a very fast clone detection stage, and it is important to note that, obviously, both a single configuration and a parametric curve require one instance of tokenization, and, therefore, the increase of the operation time for CloneWorks is not at all as high.

Finally, if clone detection takes a lot of time, as for SourcererCC, the proposed optimization has more effect. It can be seen that for a larger dataset the optimization decreases the operation time of the parametric curve search by 12.7\% (from 32 min 12 sec to 28 min 07 sec), and it can be expected that this can be even more significant for real large-scale datasets.

\subsection{BigCloneBench}

The results on the BigCloneBench are presented in Table~\ref{table:bcb}.

\begin{table}[h]
\centering
  \caption{Recall of CloneWorks for various types of clones from BigCloneBench.}
  \label{table:bcb}
  \begin{tabular}{ccccc}
    \toprule
    \textbf{Configuration} & \textbf{T1} & \textbf{T2} & \textbf{VST3} & \textbf{ST3}\\
    \midrule \vspace{0.1cm}
    \textbf{Single (77\% / 19 tokens +)} & 99.5\% & 94.9\% & 90.1\% & 37.5\%\\
    \textbf{Parametric curve} & 99.5\% & 96.4\% & 92.1\% & 59.6\%\\
  \bottomrule
  \vspace{-0.2cm}
\end{tabular}
\end{table}
\vspace{-0.2cm}

It can be seen that the parametric curve slightly increases the recall for Type-2 clones and for Very Strongly Type-3 clones, but the main change falls onto Strongly Type-3 clones, where the recall increases from 37.5\% to 59.6\%. 

%% file: sections/06-threats.tex
\section{Threats to validity}\label{threats}

The proposed approach and its optimization require a token-based clone detection tool to use the bag-of-words comparison and to support a block-level granularity. While this strongly limits the number of tools that can benefit from our approach, such tools are often used in large-scale studies for their scalability. In our research, we have limited our scope to Java as the language, SourcererCC and CloneWorks as clone detection tools, and used a set of specific projects for estimating the parameters and for the evaluation.

Also, we have used manual evaluation to estimate the necessary parameter configurations in regards to their precision, because of the large number of tested configurations. It is desirable to conduct a more thorough evaluation of the precision of the obtained configurations. We cannot be completely sure that our labeling of parameter pairs is general for all possible cases, however, two studied tools provided similar values, which might mitigate the risk. The proposed approach might be useful in cases where it is of interest to discover more clones to check them manually.

%% file: sections/07-conclusion.tex
\section{Conclusion and future work}\label{conclusion}

In this paper, we proposed and evaluated a modification to bag-of-tokens based clone detection that consists in conducting the search several times with different parameter configurations. We evaluated the approach for two clone detection tools, SourcererCC and CloneWorks, and for both of them the approach demonstrated a significant increase in the number of detected clones --- between 40.5\% and 56.6\% --- on two datasets of different size. Analyzing the results of detecting clones with CloneWorks on BigCloneBench demonstrated that the most significant increase in recall comes from Strongly Type-3 clones, which can be useful for both maintenance and inter-project clone detection. Finally, we have proposed and evaluated the optimization to the proposed approach that allowed us to decrease its operation time. This decrease depends on the operation process of the tool.

There are a lot of ways to continue this research further. It might be possible to apply EvaClone~\cite{wang2013searching} to find the best parameter configurations automatically. It is also of interest to check this approach for other types of code representation --- for example, it might be possible to employ the depth of the tree for AST-based tools, etc. Finally, we believe that the representation of clone pairs in the search space from the standpoint of their size and similarity itself may be useful for further research.